\documentclass[format=acmsmall, nonacm]{acmart}

\AtBeginDocument{%
  \providecommand\BibTeX{{%
    \normalfont B\kern-0.5em{\scshape i\kern-0.25em b}\kern-0.8em\TeX}}}

\setcopyright{none}
\copyrightyear{2020}
\acmYear{2020}

\acmConference[CSCW 2020]{Conference on Computer Supported Cooperative Work and Social Computing}{October 17--21, 2020}{Minneapolis, MN}

\usepackage{csquotes}
\usepackage{multirow}
\usepackage{multicol}
\usepackage{array}

\begin{document}

\title[Between Subjectivity and Imposition]{Between Subjectivity and Imposition: Power Dynamics in Data Annotation for Computer Vision}

\author{Milagros Miceli}
\affiliation{%
  \institution{Technische Universit{\"a}t Berlin, Weizenbaum Institut}
  \city{Berlin}
  \country{Germany}}
\email{m.miceli@tu-berlin.de}

\author{Martin Schuessler}
\affiliation{%
  \institution{Technische Universit{\"a}t Berlin, Weizenbaum Institut}
  \city{Berlin}
  \country{Germany}}
\email{schuessler@tu-berlin.de}

\author{Tianling Yang}
\affiliation{%
  \institution{Technische Universit{\"a}t Berlin, Weizenbaum Institut}
  \city{Berlin}
  \country{Germany}}
\email{tiangling.yang@tu-berlin.de}

\renewcommand{\shortauthors}{Miceli, Schuessler and Yang}

\begin{abstract}
The interpretation of data is fundamental to machine learning.  
This paper investigates practices of image data annotation
as performed in industrial contexts.
We define data annotation as a sense-making practice, where annotators assign meaning to data through the use of labels.
Previous human-centered investigations have largely focused on annotators' subjectivity as a major cause for biased labels. 
We propose a wider view on this issue: guided by constructivist grounded theory, we conducted several weeks of fieldwork at two annotation companies.
We analyzed which structures, power relations, and naturalized impositions shape the interpretation of data. 
Our results show that the work of annotators is profoundly informed by the interests, values, and priorities of other actors above their station. 
Arbitrary classifications are vertically imposed on annotators, and through them, on data. This imposition is largely naturalized.
Assigning meaning to data is often presented as a technical matter.
This paper shows it is, in fact, an exercise of power with multiple implications for individuals and society.
\end{abstract}




\maketitle

\section{Introduction}
Power imbalances related to practices of classification have long been a topic of interest for the social sciences~\cite{durkheim1963, bowker1999, bourdieu1977, bourdieu1989, bourdieu1992, mau2019}.
What is (relatively) new is that arbitrary classifications are increasingly established and stabilized through automated algorithmic systems\cite{noble2018,mau2019}.
With each system's outcome, meaning is imposed, and higher or lower social positions, chances, and disadvantages are assigned~\cite{oneil2017,barocas2016,citron2014,fourcade2013}.
These systems are often expected to minimize human intervention in decision-making and thus be neutral and value-free~\cite{seaver2019,christin2016,kitchin2017}.
However, previous research has shown that they may contain biases that lead to discrimination and exclusion in several domains such as credit~\cite{fourcade2013}, the job market~\cite{rosenblat2014}, facial recognition systems~\cite{hamidi2018,scheuerman2019,buolamwini2018}, algorithmic filtering~\cite{baker2013,noble2018}, and even advertisement~\cite{ali2019}.
Critical academic work has furthermore discussed the politics involved in data-driven systems~\cite{mager2012,dignazio2020, crawford2019} and highlighted the importance of investigating the capitalistic logics woven into them~\cite{burns2019, zuboff2019, couldry2019}.
What the enthusiasm of technologists seems to render invisible is that algorithmic systems are crafted by humans and hence laden with subjective judgments, values, and interests~\cite{gray2019,dotan2020}.
Moreover, before the smartest system is able to make predictions, humans first need to make sense of the data that feeds it~\cite{muller2019,passi2017,scheuerman2020}.
Despite its highly interpretative character, data-related work is still often believed to be neutral, ``comprising unambiguous data, and proceeding through regularized steps of analysis''~\cite{muller2019}.

The present paper investigates data annotation for computer vision based on three research questions:
How do data annotators make sense of data?
What conditions, structures, and standards shape that sense-making praxis?
Who, and at what stages of the annotation process, decides which classifications best define each data point?
We present a constructivist grounded theory~\cite{charmaz2006,muller2016,muller2014} investigation comprising several weeks of fieldwork at two annotation companies and 24 interviews with annotators, management, and computer vision practitioners.
We define data annotation as a sense-making~\cite{klein2007a} process, where actors classify data by assigning meaning to its content through the use of labels.
As we have observed, this process involves several actors and iterations and begins as clients transform their needs and expectations into annotation instructions. The sensemaking of data, so we argue, does not happen in a vacuum and cannot be analyzed independently from the context in which it is carried out.

We use Bourdieu's~\cite{bourdieu1992} concept of symbolic power, defined as the authority to impose meanings that will appear as legitimate and part of a natural order of things, as a lens to analyze the dynamics of imposition and naturalization inscribed in the classification, sorting, and labeling of data.
Previous research in the field of data annotation has largely focused on workers' individual subjectivities as a major cause for biased labels~\cite{brodley1999,hube2019,ghai2020,wauthier2011}. Conversely, our investigation introduces a power-oriented perspective and shows that hierarchical structures broadly inform the interpretation of data. Top-down meaning impositions that follow the demands of clients and the market shape data profoundly.

With this investigation, we seek to orient the discussion towards the interests and values embedded in the systems that potentially shape our individual life-chances~\cite{fourcade2013}.
Through the description of three observed annotation projects, we expose the deeply normative nature of such forms of data classification and discuss their effects on labels and datasets.
Building on this perspective, we propose the incorporation of power-aware documentations in processes of data annotation as a method to restore context. We argue that reflexive practices can improve deliberative accountability, compliance to regulations, and the explication and preservation of effective data work knowledge.
With this work, we also hope to inspire researchers 
to adopt a situated and power-aware perspective not only to investigate practices of data creation but also as a tool for reflecting power dynamics in their own research process.

\section{Related Work}
\subsection{Data Work as Human Activity}

Previous work has argued that data-driven systems are often linked to ``a technologically-inflected promise of mechanical neutrality''~\cite{gillespie2014}. However, these systems require, in many cases, the intervention of human discretion in their deployment~\cite{alkhatib2019,paakkonen2020}, and even more frequently, in their creation~\cite{muller2019,passi2017, passi2018, kitchin2017, feinberg2017,brodley1999,hube2019,finin2010,geiger2020, cheng2013}.
Moreover, critical research has argued that data-driven systems embody the personal and corporate values and interests of the people and organizations involved in their development~\cite{klinger2018,mau2019,kitchin2017,dotan2020}.
As Klinger and Svensson state, ``arguments that technology had agency on its own hide the individuals, structures, and relations in power and thus serve their interests, interests that become increasingly blurred''~\cite{klinger2018}.
A view into the power dynamics encoded in data and systems is, as we will argue, of fundamental importance, especially considering that ``the technical nature of the procedures tends to mask the presumptions that enter into the programming process, the choices that are made, and the conceivable alternatives that are ruled out''~\cite{mau2019}.

Besides technical exercise and operation, the development of data-driven products involves ``mastering forms of discretion''~\cite{passi2017} and is conditioned by the networked systems in which they are created, developed, and deployed~\cite{kitchin2017,seaver2019}.
Kitchin~\cite{kitchin2017} pinpoints various processes and factors that reveal extensive human interventions in data-driven systems, such as the translation of tasks into algorithmic models, available resources, the choice of training data, hardware and platform, the creative process of programming, and adaptation of systems to meet requirements of standards and regulations.
He further argues that algorithmic systems are subject to the purposes of their creation: ``to create value and capital; to nudge behaviour and structure preferences in a certain way; and to identify, sort and classify people''~\cite{kitchin2017}.

The examination of the provenance of data and the work practices involved in their creation is fundamental for the investigation of subjectivities and assumptions embedded in algorithmic systems.
Passi and Jackson~\cite{passi2017} propose the concept of \textit{data vision} to describe the ability to successfully work with data through an effective combination of formal knowledge and tools, and situated decisions in the face of empirical contingency.
Mastery of this interplay is essential to data analysts, which reveals ``the breadth and depth of human work'' inscribed in data ~\cite{passi2017}.

Embedded in such processes are not only individual subjectivities, but also narratives, preferences, and values related to larger socio-economic contexts~\cite{passi2018,blomberg2015,kimbell2017}: ``Numbers not only signify model performance or validity, but also embody specific technical ideals and business values.''~\cite{passi2018}.
Data practices such as the choice of training data, data capturing measurement interfaces~\cite{pine2015}, and the selection of data attributes~\cite{muller2019} as well as the design of data in an algorithmically recognizable, tractable, and analyzable way~\cite{feinberg2017,muller2019}, all indicate that data is created through human intervention~\cite{muller2019}.
Feinberg points to the ``interpretive flexibility'' and situated nature of data and considers data as a product of ``interlocking design decisions'' made by data designers~\cite{feinberg2017}.
According to Muller et al.~\cite{muller2019}, the degree of human intervention will determine how deep and fundamental subjective interpretations are inscribed in data and its analysis.

The present paper unpacks data annotation practices with a human-centered perspective.
The practices we have observed and analyzed are situated in outsourcing companies that provide annotation services for commissioning clients.
As previous work has argued, service is situated in local, cultural, and social contexts~\cite{kimbell2017} and is co-produced and co-created in the interactions between service providers and recipients~\cite{blomberg2015}.
This perspective sheds light on the situated~\cite{passi2017,elish2018} and collaborative~\cite{passi2018,feinberg2017} nature of data work, as clients and annotation teams both participate in the creation of datasets.
Scrutinizing data annotation with a service perspective further requires taking into consideration institutional structures and organizational routines~\cite{araujo2006,kimbell2017}.

Annotation tasks are, as we will argue, mainly about sensemaking~\cite{klein2007a}, i.e. framing data to make it categorizable, sortable, and interpretable.
Previous work in this space has largely focused on individual preconceptions, considering annotators' subjectivities to be a major source for labeling bias~\cite{brodley1999, ghai2020,wauthier2011,hube2019}.
However, other researchers (we among them) explore factors beyond individual subjectivities that influence workers and labels, such as loosely-defined annotation guidelines and annotation context~\cite{finin2010}, the choice of annotation styles~\cite{cheng2013}, and the interference between items in the same data batch~\cite{zhuang2015}.
In a thorough investigation into annotation practices in academic research, Geiger et al.~\cite{geiger2020} draw attention to the background of annotators, formal definitions and qualifications, training, pre-screening for crowdwork platforms, and inter-rater reliability processes.
The authors consider these factors to be likely to influence the annotations and advocate for their documentation.

With the present paper, we join the discussion around subjectivity in data annotation.
By examining the processes and contexts that shape this line of work, we argue that subjectivity can also be shaped by power structures that enable the imposition of meanings and classifications.

\subsection{Data, Classification, Power}

Practices related to classifying and naming constitute the core of data annotation work.
As Bowker and Star~\cite{bowker1999} have most prominently argued, classifications represent subjective social and technical choices that have significant yet usually hidden or blurry ethical and political implications~\cite{zerubavel1993}.
Classification practices are constructed and, at the same time, construct the social reality we perceive and live in~\cite{bourdieu1989}.
Therefore, they are also culturally and historically specific~\cite{hanna2020}.
Adopting a critical position to examine these practices is essential because, as Durkheim and Mauss argue,
``every classification implies a hierarchical order for which neither the tangible world nor our mind gives us the model.
We therefore have reason to ask where it was found''~\cite{durkheim1963}.

Humans collect, label, process, and analyze data in the usually invisible context of a design plan, where what is considered data~\cite{boyd2012, pine2015}, and how data is to be classified \cite{bowker1999} is established.
``A dataset is a worldview'', as Davis~\cite{davis2020} wonderfully puts it.
Accordingly, it can never be objective nor exhaustive because,
``it encompasses the worldview of the labelers, whether they labeled the data manually, unknowingly, or through a third party service like Mechanical Turk, which comes with its own demographic biases.
It encompasses the worldview of the built-in taxonomies created by the organizers, which in many cases are corporations whose motives are directly incompatible with a high quality of life.''~\cite{davis2020}.
Furthermore, decisions about what information to collect and how to measure and interpret data define possibilities for action by making certain aspects of the social world visible – thus measurable – while excluding other aspects~\cite{pine2015,dignazio2020}.
Data-related decisions are infrastructural decisions \cite{pine2015,bowker1999} as they ``exercise covert political power by bringing certain things into spreadsheets and data infrastructures, and thus into management and policy''~\cite{pine2015}.
This way, datasets are powerful technologies~\cite{bowker1999} that bring into existence what they contain, and render invisible what they exclude.
As Bowker argues, ``the database itself will ultimately shape the world in its image: it will be performative.''~\cite{bowker2000b}.

The performative character of datasets, that is, the power of creating reality through inclusion and exclusion, relates to Pierre Bourdieu’s theorization of \emph{symbolic power}.
Symbolic power is the authority to sort social reality by separating groups, classifying, and naming them \cite{bourdieu1992, bourdieu1985}.
Every act of classification is an attempt to impose a specific reading of the social world over other possible interpretations~\cite{mau2019}.
Thus, symbolic power is not merely a matter of naming or describing social reality but a way of ``making the world through utterance''~\cite{bourdieu1992}.
The \emph{power} aspect here relates to the authority to lend legitimacy to certain definitions while delegitimizing others.
This authority is unevenly distributed and correlates with the possession of economic, cultural, and social capital~\cite{bourdieu1989}.

According to Bourdieu~\cite{bourdieu1992}, dominant worldviews find their origin in arbitrary classifications that serve to legitimize and perpetuate power asymmetries, by making seem natural what is in fact political:
``Every established order tends to produce (to very different degrees and with very different means) the naturalization of its own arbitrariness''~\cite{bourdieu1977}.
The systems of meaning created through acts of symbolic power are arbitrary because they are not deducted from any natural principle but subject to the interests and values of those in a dominant position at a given place and time in history~\cite{bourdieu1990}.
A combination of \textit{recognition} and \textit{misrecognition} is necessary to guarantee the efficacy of arbitrary classifications~\cite{bourdieu1977}:
the authority to impose classifications must be \textit{recognized as legitimate}, for the imposition to actually be \textit{misrecognized in its arbitrariness} and be perceived as natural.
This process of \emph{naturalization} allows arbitrary ways of sorting the social world to become so deeply ingrained that people come to accept them as natural and indisputable.
As argued by D’Ignazio and Klein~\cite{dignazio2020}, ``once a [classification] system is in place, it becomes naturalized as `the way things are'{''}.
Thus, the worldviews imposed through symbolic power are rendered less and less visible in their arbitrariness, until disappearing into the realms of what is considered common sense.
As we will argue, the interplay between recognition of authority and naturalization of arbitrary classifications decisively shapes annotations and data.

Previous investigations have related discriminatory or exclusionary outputs of data-driven systems to symbolic power:
Mau argues that ``advancing digitalization and the growing importance of Big Data have led to the rapid rise of algorithms as the primary instruments of nomination power''~\cite{mau2019}. Here, nomination refers to the authority to name and classify.
The author argues the ubiquity of an algorithmic authority, connected to a wide range of procedures participating in the reinforcement of social classifications.
Crawford and Paglen~\cite{crawford2019} discuss the politics involved in training sets for image classification.
The authors expose the power dynamics implicit in the interpretation of images as it constitutes ``a form of politics, filled with questions about who gets to decide what images mean and what kinds of social and political work those representations perform''~\cite{crawford2019}.
Even if not directly referenced to Bourdieu, Crawford and Paglen's conclusion closely relates to what the French sociologist has described as the ``social magic''~\cite{bourdieu1992} of creating reality through naming and classifying:
``There’s a kind of sorcery that goes into the creation of categories.
To create a category or to name things is to divide an almost infinitely complex universe into separate phenomena.
To impose order onto an undifferentiated mass, to ascribe phenomena to a category—that is, to name a thing—is in turn a means of reifying the existence of that category.''~\cite{crawford2019}

Investigating data as a human-influenced entity~\cite{muller2019} informed by power asymmetries~\cite{barabas2020} means understanding both data and power relationally.
Data exists as such through human intervention~\cite{muller2019} because, as we have seen, ``raw data is an oxymoron''~\cite{gitelman2013}.
Similarly, Bourdieu~\cite{bourdieu1985} offers a relational view of power as enacted in the interaction among actors as well as between actors and field.
In the discussion section, we will analyze the relation between annotators, data, and corporate structures.
The symbolic power construct will then offer a valuable contribution to the discussion of assumptions encoded in datasets that reflect the naturalization of practices and meanings~\cite{cronin1996,bourdieu1977}.

\section{Method}
This investigation was guided by three research questions:

\textbf{RQ1:} How do data annotators make sense of data?

\textbf{RQ2:} What conditions, structures, and standards shape that sense-making praxis?

\textbf{RQ3:} Who, and at what stages of the annotation process, decides which classifications best define each data point?

We followed a constructivist variation of grounded theory methodology (GTM)~\cite{charmaz2006,muller2014,muller2016}.
The central premise of constructivist GTM is that neither data nor theories are discovered, but are formed by the researcher's interactions with the field and its participants~\cite{thornberg2012}. This method provided tools to systematically reflect on our position, subjectivity, and interpretative work during fieldwork and at the coding stage.

Data was obtained through participatory observation (with varying degrees of involvement) and qualitative interviewing (in-depth and expert interviews).
Fieldwork was approached exploratorily, guided by sensitizing concepts~\cite{charmaz2006}.
They helped to organize the complex stimuli in the field without acting as hypotheses or preconceptions.
Phases of data collection and analysis were intertwined.
Observations and interviews informed one another:
while ideas emerging from the observations served to identify areas of inquiry for the interviews and even possible relevant interview partners, statements from the interviews pointed many times at interesting actors, tasks, or processes needing to be more attentively observed.
Through constant comparison~\cite{glaser1998}, we were able to identify differences and similarities between procedures and sites.

\subsection{Data Collection}
\subsubsection{Participatory Observations}
Part of the value of open-ended observations guided by GTM is the opportunity to see the field inductively and allowing themes to emerge from the research process and the data collected.
However, once in the field, researchers must somehow organize the complex stimuli experienced so that observing becomes and remains manageable because it is certainly not possible to observe all details of all situations.
At this point, sensitizing concepts come into play to orient fieldwork~\cite{charmaz2006}.
Sensitizing concepts in this investigation include loosely defined notions such as ``impact sourcing'', ``subjectivity'', ``quality assurance'', ``training'', and ``company's structure'', which provided some initial direction to guide the observation during data gathering.

Fieldwork was conducted at two data annotation companies.
At both locations, the level of involvement regarding observations varied from shadowing to active participant observations~\cite{morike2019}.
At both annotation companies, fieldwork was allowed to commence after a representative of the company and the researcher on the field signed non-disclosure agreements (NDA) and respectively consented for participating in the present study.
Therefore, we are restrained from disclosing or using confidential information in this paper, particularly concerning the companies' clients.

\subsubsection{Qualitative Interviews}
Part of the fieldwork conducted consisted of intensively interviewing annotators and management.
All interview partners were allowed to choose their code names or were anonymized post-hoc to preserve the identity of related informants.

Interviews with management in additional annotation companies were framed as expert interviews.
While in-depth interviews aim at studying the informant's practices and perceptions, ``the purpose of the expert interview is to obtain additional unknown or reliable information, authoritative opinions, serious and professional assessments on the research topic''~\cite{libakova2015}.
The sampled interview partners were considered experts because they provided unique insights into the structures and processes within their companies and the overall market (see table \ref{table:informants} and section \ref{section:sample}, Sample, for a detailed list of informants).

\subsection{Sample}\label{section:sample}
Four sources of information were exhaustively explored:
we started with two impact sourcing companies dedicated to data annotation located in Buenos Aires, Argentina (S1) and Sofia, Bulgaria (S2).
Impact sourcing refers to a branch of the outsourcing industry employing workers from poor and vulnerable populations to provide information-based services at very competitive prices.
We chose annotation companies with rather traditional management structures over crowdsourcing platforms where hierarchies appear not as evident
We assumed that clear hierarchical structures would make it easier to trace back labeling decisions and structures to real people.
We also had the preconception that tensions related to exercising power would be more prominent with workers from vulnerable populations.
Field access was another reason for our choice.
Impact sourcing companies responded most openly to our proposed ethnographic research.

While conducting fieldwork in S2, we decided to look closer into the translation of clients' needs into annotation tasks and quality standards.
Consequently, we also interviewed management employees in three similar yet larger annotation companies (S3) and engineers with a computer vision company using annotated training sets in Berlin, Germany (S4).

\begin{table}
\caption{Overview of Informants and Fieldwork Sites}
\label{table:informants}
\tiny
\def\arraystretch{1.2}
\begin{tabular}{p{2.1cm}p{2.1cm}p{3.5cm}p{1.2cm}p{3.1cm}}
    \hline
    \hline
    & \textbf{Interview method} & \textbf{Medium and Language} & \textbf{Code name} & \textbf{Role} \\
    \hline
    \multirow{6}{2.1cm}{\textbf{S1: FIELDWORK} (Annotation company in Buenos Aires, Argentina)} &
    \multirow{6}{2.1cm}{Qual. in-depth interview} &
    \multirow{6}{2.6cm}{Face to face; Spanish (Native)}
    & Sole & Team leader \\ \cline{4-5}
    &&& Elisabeth & Annotator, reviewer \\ \cline{4-5}
    &&& Noah & Annotator, tech leader \\ \cline{4-5}
    &&& Natalia & Project manager \\ \cline{4-5}
    &&& Paula & Founder \\ \cline{4-5}
    &&& Nati & QA analyst \\
    \hline
    
    \multirow{11}{2.1cm}{\textbf{S2: FIELDWORK} (Annotation company in Sofia, Bulgaria)} &
    Qual. expert interview & Skype, Face-to-face; English (Proficient) & Eva & Founder \\
    & Qual. in-depth interview & Face-to-face; English (Proficient) & & \\
        \cline{2-5}
        
        &\multirow{2}{2.1cm}{Qual. expert interview}
        & Face to face; English (Proficient) & Anna & Intern in charge of impact assessment \\ \cline{3-5}
        && Face to face; English (Low-intermediate) & Ali & Project manager, reviewer \\
        \cline{2-5}
        
        &\multirow{8}{2.1cm}{Qual. in-depth interview}
        & Face to face; English (Low-Intermediate) & Savel & Annotator \\ \cline{3-5}
        && Face to face; English (Upper-Intermediate) & Diana & Annotator \\ \cline{3-5}
        && Face to face; English (Low-Intermediate) with occasional translation by another informant
            & Hiva & Annotator \\ \cline{3-5}
        
        &&\multirow{3}{3.5cm}{Face to face; English (Intermediate)}
        & Mahmud & Annotator \\ \cline{4-5}
        &&& Mariam & Annotator \\ \cline{4-5}
        &&& Martin & Annotator \\ \cline{3-5}
        && Face to face; English (Advanced) & Sarah & Annotator \\ \cline{3-5}
        && Face to face; another informant translated into English (Advanced) & Muzhgan & Annotator \\
        \hline
        
    \multirow{3}{2.1cm}{\textbf{S3: EXPERTS} (Managers in large annotation companies)} &
    \multirow{3}{2.1cm}{Qual. expert interview} & 
    \multirow{2}{3.5cm}{Zoom, English (Proficient)}
    & Jeff & General manager in annotation company in Iraq \\ \cline{4-5}
    &&& Gina & Program manager in annotation company in Iraq \\ \cline{3-5}
    && Zoom, English (Native) & Adam & Country manager in annotation company in Kenya \\ \cline{3-5}
    && Zoom, English (Advanced) & Robert & Director in annotation company in India \\
    \hline
    
    \multirow{4}{2.1cm}{\textbf{S4: PRACTITIONERS} (Computer vision company in Berlin, Germany)} &
    \multirow{4}{2.1cm}{Qual. in-depth interview}
    & Face to face; English (Proficient) & Ines & Project manager, data protection officer \\ \cline{3-5}
    && Face to face; English (Advanced) & Dani & Product manager \\ \cline{3-5}
    && Face to face; English (Advanced) German (Native) & Michael & Computer vision engineer \\ \cline{3-5}
    && Face to face; English (Proficient) & Dean & Research scientist, lead engineer \\
    \hline
    \hline
        
\end{tabular}
\end{table}

\subsubsection{S1: The Annotation Company in Buenos Aires}
At the time of this investigation in June 2019, S1 is a medium-sized enterprise centrally located in Buenos Aires and dedicated to data-related microwork. 
The company has further branches in Uruguay and Colombia.  
The Buenos Aires office occupies a whole floor with large common work areas. 
This location employs around 200 data workers, mainly young people living in very poor neighborhoods or slums in and around Buenos Aires.
The companyäs employment strategy is a conscious decision as part of its impact sourcing mission.
At S1, workers are divided into four teams.
Each team includes a project manager and several team leaders and tech leaders.
Annotators perform their tasks in-house and assume mainly two roles:
\textit{creators}, doing the actual labeling work, or \textit{reviewers}, who confirm or correct annotations.
Besides annotations for visual data, the company also conducts content moderation and software testing projects.
Most of the clients are large local or regional companies, including media, oil, and technology corporations.
At the time of this investigation (June 2019), S1 had just started to expand to Brazil and other international markets, which resulted in the need to train their workers in Portuguese and English.

One particularity of S1 is that they provide workers with a steady part- or full-time salary and benefits.
This form of employment contrasts with the widespread contractor-based model in data annotation.
Even so, annotators at S1 received USD1.70 per hour, the minimum legal wage in Argentina at the time of this investigation.
These salaries left workers way below the poverty line in a country that accumulated around 53.8\% annual inflation in 2019.
Low salaries are not the only downside perceived by workers: informants also complained about the fixed work shifts and the impossibility to work remotely, as the company does not allow its workers to take laptops or any other equipment home.

The interviews at the Argentine company were conducted in Spanish, the mother tongue of both interviewer and informants.
Interview transcripts were coded and interpreted without translating them by the first and third authors, native and intermediate Spanish speakers, respectively.
Coding without translation was done to preserve the original meaning of the statements.
The quotations in this paper were translated upon completion of the analysis.

\subsubsection{S2: The Annotation Company in Sofia}
S2 is a small annotation company in the center of Sofia, Bulgaria. 
The company occupies a relatively small office.
Work at this location can be quite chaotic, with workers coming and going to receive paychecks or instructions for new projects.
The company focuses on the annotation of visual data, especially image segmentation and labeling.
The visual data involves various types of images, including medical residue, food, and satellite imagery.
The company's clients are mostly located in Europe and North America.
At the time of this investigation in July 2019, ten active projects were handled by three employees in salaried positions and a pool of around 60 freelance contractors.
As an impact sourcing company, S2 is committed to fair payment and works exclusively with refugees and migrants from the Middle East.
The company also favor female workers among them.
Contractors mostly work remotely with their own or company-provided laptops, with flexible hours.
They are paid per picture and, sometimes, per annotation.
Payment varies according to the project and the level of difficulty.
Most informants were satisfied with the remuneration and flexible conditions.
However, many of them expressed the desire to have more stability and continuity of work and income.

All interviews at this location were conducted in English.
Most annotators had low to medium English skills, which represented a significant difficulty for the conduction of interviews.
For example, some informants over-simplified their statements and were often not able to provide in-depth answers.
The language barrier could not have been foreseen or mitigated, as the founder, whose English skills are impeccable, had assured us a selection of interview partners with similar language skills. 
The misunderstanding probably originated in the fact that all proposed informants were indeed able to understand English at a level that was sufficient to perform their work. 
It was, however, not enough for them to easily tell their stories.
The language barrier required improvisation on researchers end, including the simplification of questions and the introduction of walk-through interviews~\cite{morike2019}, allowing informants to show procedures directly while reducing language requirements (see table \ref{table:informants} for more details).

\subsubsection{S3: The Experts}
In grounded theory investigations, decisions regarding theoretical saturation often happen simultaneously with the gathering of data, forcing researchers to make quick decisions on whether the collection of further or different data is necessary.
While conducting fieldwork in Bulgaria, the idea emerged that expert interviews with management in other, more prominent impact sourcing companies could provide further insights about the translation of clients' needs into actual annotation tasks, standards, and quality assurance (QA). Through this form of inquiry, we additionally sought to frame some of the fieldwork observations. 

Three expert interviews were conducted:
\textit{Jeff and Gina} are, respectively, general and program manager with a microwork company based in Iraq.
Jeff is also in charge of training future workers on data annotation.
The company had initially been founded by a worldwide organization dedicated to humanitarian aid and quickly became a for-profit impact sourcing company. Jeff and Gina were interviewed simultaneously.
\textit{Adam} is the general manager at the Kenyan branch of an impact sourcing company with many hubs for data annotation throughout Asia and Africa.
\textit{Robert} is based in India and works as a director of machine learning with one of the oldest impact sourcing companies dedicated to data annotation.
The company has many branches in different Asian countries.

The informants are identified through code names.
The names of their companies remain anonymous.

\subsubsection{S4: The Practitioners}
The demands, rules, and processes of clients represented a recurring topic of the interviews conducted within data labeling companies.
It seemed that managerial roles within labeling companies implied the ability to mediate and translate the client's requirements into factual tasks for workers.

How do such requirements originate? On whose needs are they based?
To start exploring these questions, we decided to briefly investigate companies ordering and deploying labeled datasets for their machine learning products.
A visit to a computer vision company based in Berlin was then arranged and carried out.
Four relevant actors were interviewed in-depth at this location:
a project manager, the data protection officer, the lead engineer, and a data engineer.

While this company is not a direct client of S1, S2, or S3, it does commissions and utilizes labeled images for its main product.

\begin{table}[h]
\caption{Table of core phenomenon, axial categories, open codes, and explanatory memos.}
\label{table:codescategoriesmemos}
\tiny
\def\arraystretch{1.2}
\begin{tabular}{p{0.3cm}p{1.6cm}p{0.8cm}p{1.1cm}p{8cm}}
\cline{2-5}
\cline{2-5}
\parbox[t]{2mm}{\multirow{3}[130]{*}{\rotatebox[origin=c]{90}{\textbf{IMPOSITION OF MEANING}}}}
& \multicolumn{1}{c}{\textbf{Axial Categories}}
& \multicolumn{2}{c}{\textbf{Open Codes}}
& \multicolumn{1}{c}{\textbf{Memos}} \\
\cline{2-5}
&\multirow{5}{1.6cm}{{CLASSIFICATION AS POWER EXERCISE}}
    & \multicolumn{2}{l}{Briefing}
    & Information of labeling instruction to labelers.
      Communication of client's wishes and expectations.
      Communication chain from client to labelers.
        \\ \cline{3-5}
        
&    & \multicolumn{2}{l}{Struggle over meaning}
    & Struggle over the meaning of things.
      Power struggles to name things.
      Also moments of subversion from labelers.
        \\ \cline{3-5}
        
&    & \multicolumn{2}{l}{Imposition}
    & One-way, top-down imposition of meaning during team meetings.
      Imposition of client’s desires and/or views in view of discrepancies.
        \\ \cline{3-5}
        
&    & \multicolumn{2}{l}{Team Agreement}
    & Democratic alignment of concepts and opinions within the team.
      Teamwork to reach an agreement on how to name things.
        \\ \cline{3-5}
        
&    & \multicolumn{2}{l}{Layering}
    & Nomination instances within annotation companies.
      Actors deciding over the interpretation of data at different stages of the process.  
        \\ \cline{2-5}
 
&\multirow{4}{1.6cm}{{LABELING OF DATA}} 
    & \multicolumn{2}{l}{Tools}
    & Different tools to perform tasks of data annotation,
      where they come from and how they may represent a constraint for the work.
        \\ \cline{3-5}
        
&    & \multicolumn{2}{l}{Agency}            
    & Room for agency while performing labeling tasks;
      agency here refers to the possession of resources to achieve desired results.
        \\ \cline{3-5}
&    & \multicolumn{2}{l}{Constraints}       
    & Things that could count as a constraint for subjectivity when performing tasks of data annotation.
        \\ \cline{3-5}
        
&    & \multicolumn{2}{l}{Standardization}   
    & How labeling is standardized. 
      Efforts from company or client to standardize labeling tasks.
        \\ \cline{2-5}
                          
&\multirow{2}{1.6cm}{{REFLEXIVITY ON WORK IMPACT}}
    & \multicolumn{2}{l}{Visions of future} 
    & How workers imagine the future in relation to the tasks they perform.
        \\ \cline{3-5}
        
&    & \multicolumn{2}{l}{Tech}              
    & Visions of impact of technology /digitalization/AI on society.
      Impact of their work on society.
        \\ \cline{2-5}

&\multirow{6}{1.6cm}{{IMPACT SOURCING}}
    & \multicolumn{2}{l}{Training}                  
    & Training received as part of the impact sourcing model.
      Training that could be helpful for future jobs (languages, software, etc).
        \\ \cline{3-5}
        
&    & \multicolumn{2}{l}{Chance}                    
    & Chances to learn, to work in the desired field. 
      Chances related to impact sourcing companies.
      Opportunities offered by companies to their employees.
        \\ \cline{3-5}
        
&    & \multicolumn{2}{l}{Impact on lives}           
    & Impact of job on worker’s lives.
      What this job means for them and how their lives have changed with this job.
        \\ \cline{3-5}
        
&    & \multicolumn{2}{l}{Closeness to management}   
    & Indicators for flat hierarchies.
      Accessibility to management.
      Possibility to talk directly and honestly to management.
        \\ \cline{3-5}
        
&    & \multicolumn{2}{l}{Recruting}                 
    & How the interviewee was recruited to work in the company.
      How she/he got to work there.
        \\ \cline{3-5}
        
&    & \multicolumn{2}{l}{Mobility chances} 
    & Chances to grow and/or be promoted within the company.
        \\ \cline{2-5}

&\multirow{3}{1.6cm}{{BIAS}}
    & \multicolumn{2}{l}{Misunderstanding}  
    & When the interviewer asks about biases and the interview partner
      offers an answer showing they have misunderstood the question.
        \\ \cline{3-5}
        
&    & \multicolumn{2}{l}{Unawareness}       
    & Not knowing what the concept of bias refers to.
      Not being aware of biases as a hazard related to their tasks.
        \\ \cline{3-5}
        
&    & \multicolumn{2}{l}{Not bias related}  
    & Claim that biases are not relevant for the type of projects
      they handle within the company
        \\ \cline{2-5}

&\multirow{5}{1.6cm}{{CAPITALISTIC LOGISTICS}}

    & \multirow{5}{0.8cm}{Company structure}
        & Speed 
        & Optimization of processes,
          so that they are faster and the client is satisfied.
            \\ \cline{4-5}
&        && QA           
        &  Quality Assurance Processes. 
           Especially QA as a selling argument for clients.
           Control as a selling point.
            \\ \cline{4-5}
&        && Productivity
        &  Processes related to increasing or controlling productivity,
           making workers produce more.
            \\ \cline{4-5}
&        && Flexibility
        &  Flexibility in working time, work place;
           not as a fixed/ regular employee; work with children etc.
            \\ \cline{4-5}
&        && Roles        
        &  The division of roles and tasks in the work/ in the company.
        \\ \cline{3-5}

&    & \multicolumn{2}{l}{Market's logics}   
    & Things that are done in a certain way to go according to 
      the demands of the market.
        \\ \cline{3-5}
        
&    & \multicolumn{2}{l}{Worker's struggle} 
    & Workers asking for better conditions/benefits.
      Expressions of disagreement with aspects of the working conditions.
        \\ \cline{3-5}
        
&    & \multicolumn{2}{l}{Control}           
    & Control mechanisms.
      Control of results.
      Control of employees.
        \\ \cline{3-5}
        
&    & \multicolumn{2}{l}{Clients}           
    & All things clients.
      Communication with clients, desires of the clients, relation to clients.
      Client as king.
        \\ \cline{2-5}
        
&\multirow{4}{1.6cm}{{PERSONAL SITUATION}}
    & \multicolumn{2}{l}{Plans}  
    & Plans for the future at a personal level. 
      Hopes and dreams.
        \\ \cline{3-5}
        
&    & \multicolumn{2}{l}{Vulnerability}       
    & Related to the vulnerable background of workers. 
      Personal struggle/ difficulties.
        \\ \cline{3-5}
    
&    & \multicolumn{2}{l}{Previous work experience}  
    & What workers did before becoming labelers.
        \\ \cline{3-5}
        
&    & \multicolumn{2}{l}{Education}  
    & Related to workers' background. 
      Achieved academic level. 
      Plans for further education.
        \\ \cline{2-5}
        \cline{2-5}
\end{tabular}
\end{table}

\subsection{Data Analysis}
The resulting 24 interviews were transcribed.
Transcriptions were integrated with several pages of field notes and various documents such as specific instructions provided by clients with labeling requirements, metrics for quality assurance, and impact assessments.
We followed the grounded theory coding system~\cite{charmaz2006} for the interpretation of data:
Phases of \emph{open, axial}, and \emph{selective coding} were systematically applied.

By the end of the \emph{open coding} phase, a set of 28 codes had emerged.
The process of \emph{axial coding} followed.
We applied a set of premises~\cite{corbin2015} to make links between categories visible.
The material was then meticulously coded using the renewed set of axial categories. As part of this process, we iteratively returned to the material to look for additional evidence and to test and revise the emergent understanding.
This analysis led to a core set of seven axial codes (see Table \ref{table:codescategoriesmemos}).
For the \emph{selective coding}, the final step of the analysis, we combined several axial codes to the core phenomenon \emph{``imposition of meaning''}.
Selective coding indicates deliberate interpretive choices by the researchers. 
Making such choices explicit during the analysis process is fundamental in constructivist grounded theory~\cite{charmaz2006}.

\begin{figure}[h]
    \centering
    \includegraphics{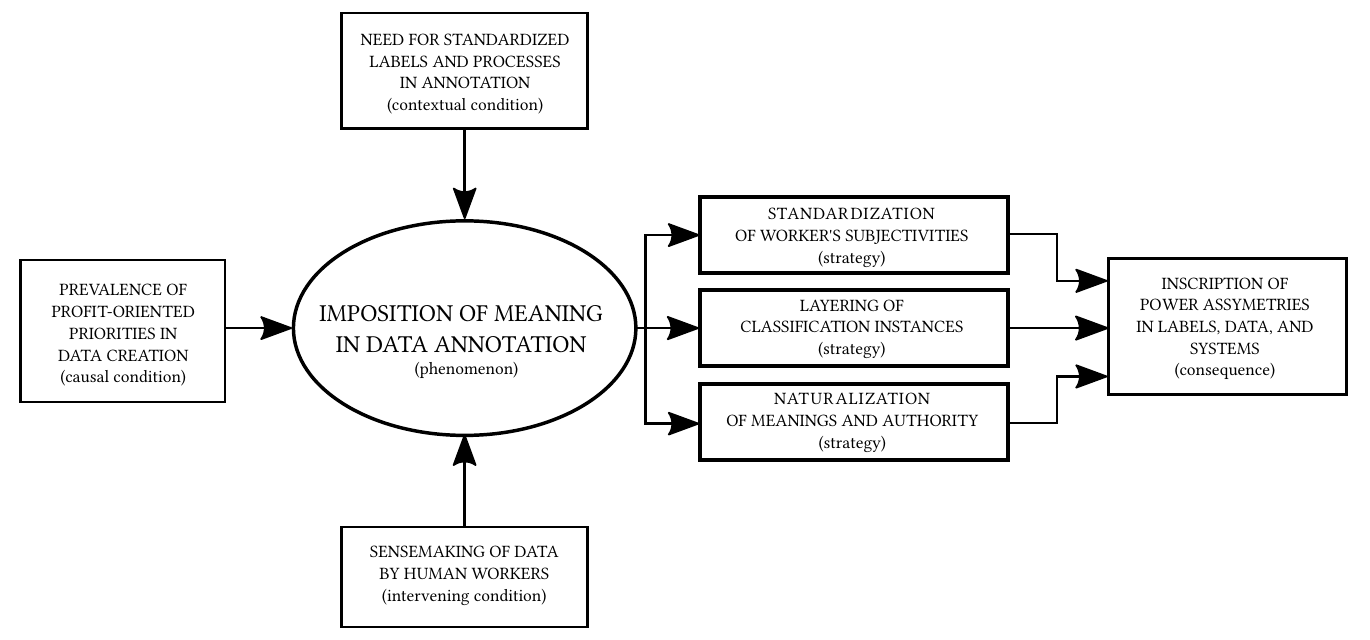}
    \caption{Paradigm model resulting from the process of selective coding. It depicts the top-down allocation of meanings, its stabilization through annotation practices, and its effects on data (derived from the Grounded Theory Paradigm Model, by Corbin and Strauss). \cite{corbin2015}}
    \label{fig:paradigm}
\end{figure}

Finally, we connected salient codes and categories to the core phenomenon as causal conditions, context, intervening conditions, action/interactional strategies, or consequences~\cite{corbin2015} (see Figure \ref{fig:paradigm} ``Paradigm Model'').

\section{Findings}
The annotation of visual data consists of a set of practices aiming at interpreting the content of images and assigning labels according to that interpretation.
The observed work practices involve mainly two tasks:
labeling and segmenting.
Segmenting, formally called semantic segmentation, refers to the separation of objects within an image, thus classifying them as belonging to different kinds.
Labeling is mainly about giving a name to each of the objects that were previously classified as different from each other.
Sometimes, labeling also includes the assignment of keywords and attributes.
Those attributes fill the performed classifications with meaning by putting in words what constitutes each class.

To illustrate our findings, we describe three of the observed annotation projects, that were particularly relevant to our research questions.
Several of the practices and tensions described in these cases remained consistent across projects and even companies.
Finally, we report four salient observations that emerged from the collected data as part of the coding process.

\subsection{Project 1: Drawing Polygons} \label{sec:proj1}
This project, conducted by S2 in Bulgaria, consisted of analyzing, marking, and labeling pictures of vehicles for a Spanish client.
The client had provided several image collections, each containing photographs of damaged car exteriors.
The source of the images and the exact purpose of the dataset were unclear for the Bulgarian team.
Only Eva, the founder of the annotation company, was capable of sharing some vague information about the client and the planned product:
\begin{displayquote}
``I think it's a company working for insurance companies.
So, they are providing insurance companies with a tool or a service I believe that's going to be in the form of an app that their users, who are using the insurance or maybe car rental companies and so on, can use in order to report damages.
And so, these damages can be processed very quickly and identify them automatically.
I think this is the final goal.
I believe they are in the very early stage still.
They are still trying to gather enough photos and train enough, use enough data to train their models.''
\end{displayquote}
Eva was in charge of client communications and the final quality control for every project at S2.
Ali, an annotator who generally acted as mediator between Eva and the team, worked on the project as well.
Besides regular annotation tasks, Ali was in charge of selecting the annotators for this project, briefing them with the instructions, and answering questions.
For this purpose, he maintained a project-specific Slack channel.
Daily, he monitored the progress made by every labeler and reviewed the annotated pictures.
Despite his prominent role, Ali had no information about the planned product or the purpose of the annotations.
Lack of information and general unawareness of the machine learning pipeline was very common among annotators at S2 and, to a lesser extent, at S1 in Argentina.
Eva agreed with this observation and added:
\begin{displayquote}
``I think that in many cases it's too difficult for a lot people to imagine what's the data they're working on for.''
\end{displayquote}
Besides Eva, none of the annotators we interviewed in S1 could relate the terms ``machine learning'' or ``artificial intelligence'' to their work.
Ali did not inquire about further details beyond the specific instructions for the ``car accidents project'' because the instruction sent by clients normally provided ``all we need'' to complete annotation tasks:  
\begin{displayquote}
I: ``But why does the client need all these pictures annotated like this? Do you know?''\\
B: ``No.
But I think ...
I am not sure, because I don't ask about this.''
\end{displayquote}
In this case, the client had sent a PDF document containing step-by-step instructions and example pictures.
Moreover, the client had provided the platform where the segmentation and annotation tasks were to be performed.
The platform had been specially developed for this purpose and tailored to the client's needs.

The first task for the annotators was to select the part of the vehicle that appeared damaged from a sidebar containing different classes (e.g., door, tire, hood).
After that, they drew a polygon around the damaged area.
The drawing was very time-consuming, and Ali seemed to pay special attention to the correct demarcation of the damaged areas.
After drawing the polygon, they would classify the type of damage and its severity.
Unfortunately, the company commissioning these annotations requested that no further details about the specific commands and labels are shared in this investigation as the company considers them one of their strategic advantage.

Apart from Eva and Ali, five annotators working remotely completed the project team.
For the general briefing and the project kick-off, they were summoned to the office.
Eva explained the client's instructions in English and showed some examples of the pictures and the procedure.
Ali translated into Arabic for annotators with low English skills.
Afterward, each annotator sat at one of the work stations in the office and tested the task while Ali walked around observing how annotators performed, answering questions, and continuously commenting on how easy the work was.
For the duration of the project, annotators working remotely would resolve questions with Ali via Slack.
Occasionally, if Ali was not satisfied with the quality of the polygons, he would summon the annotators to the office and work with them for a few hours.
The same procedure was followed in cases of visible annotation inconsistencies among different workers.
Eva highlighted the importance of these ``alignment meetings'' to ensure the uniformity of the labels through the standardization of workers' subjectivities:
\begin{displayquote}
``Normally, issues in data labeling do not come so much from being lazy or not doing your work that well.
They come from a lack of understanding of the specific requirements for the task or maybe different interpretations because a lot of the things ...
Two people can interpret differently so it's very important to share consistency and like having everyone understand the images or the data in the same way [...].
But because a lot of these tasks are not that straightforward, it's just not ...
It's not just choosing A or B.
It's more like okay for example I have this car, where do I track the exact scratch or deformation?
What kind of a level is it?
Like, it's a little bit more complicated and that's why it's better to invest in the human capability and let's say the standardization of everyone's understanding.''
\end{displayquote}

\subsection{Project 2: Building Categories} \label{sec:proj2}
This project was conducted at S1, the Argentine annotation company. 
It constituted a test for the acquisition of an important client, namely a sizable local corporation.
The potential client had simultaneously outsourced the project with different annotation companies, planning to sign a contract with the best performing team.

We find this project to be particularly interesting as it constitutes an exception to the usual procedure of labeling data according to categories instructed by clients.
In this case, the annotators were in charge of developing a classification system for the annotations.
Concretely, the task consisted of analyzing camera footage, counting, and classifying vehicles driving in a gas station.
The annotators were in charge of coming up with logical, mutually exclusive categories for the labeling.

Three annotators, a reviewer, a team leader, and a quality assurance (QA) analyst sat together to analyze the first, 60-minutes-long video.
They started by counting \emph{all} vehicles driving in the gas station.
After a few minutes, some analysts lost track and claimed they did not expect ``just counting'' to be so complicated.
To simplify the task, the team leader suggested establishing categories first, so that each annotator could focus on counting only one category.
They promptly agreed on five categories, namely cars, buses, trucks, motorcycles, and vans.
While counting, new categories such as pick-ups, SUVs, and semi-trucks were suggested by annotators, approved by the team leader and the QA analyst, and finally added to the list.
Several questions arose:
Can SUVs be considered cars?
Do ambulances and police cars constitute categories for themselves?
Furthermore, several team members expressed being worried about not knowing the client's exact expectations.
``We are not really used to this kind of ambiguity'' reviewer Elisabeth said.
She also shared an experience from a former project, where inconsistencies between the interpretations of client and annotators had arisen, even though the client had provided clear instructions for the annotations.
On that occasion, Elisabeth had been entirely sure that her interpretation was right until the client corrected her work:
``and you think you're doing everything right until the client comes and says, `No, that’s all wrong!'{''}
The client's correction had led Elisabeth to the conclusion that ``I had been wrong all along.
It put us [the team] back on track.''

As for the ``gas station project'', Nati, the QA analyst, announced to the team that, despite the freedom offered by the project, they would proceed "as usual" to resolve questions and, most importantly, to assess the correctness of allocated labels.
Upon request of the interviewer, reviewer Elisabeth described the usual process in detail:
\begin{displayquote}
``Whenever I cannot resolve the questions annotators bring to me, I ask the leader.
If the leader cannot solve them either, we ask QA.
Otherwise, they ask the contact person at the client's company.''\\
Interviewer: ``So, the client has the final say?''\\
Elisabeth: ``Yes.
And the client surely has their hierarchies to discuss a solution as well.''
\end{displayquote}
Despite the room offered to the team by the ``gas station project'' to shape data according to their own judgment, the client's figure seemed to be tacitly present at all times to orient annotators' subjectivities.
QA analyst Nati summarized this observation most clearly:
\begin{displayquote}
``We try to guess what the client would value the most, what will interest them, trying to put ourselves in their shoes, thinking, imagining [the client] wants this or that.''
\end{displayquote}
In her QA analyst role, Nati also paid special attention to optimizing the time needed to annotate each video.
Having one annotator counting only one category significantly reduced task completion time but raised important questions about quality control and cost optimization, as Nati pointed out:
\begin{displayquote}
``How are we going to check for accuracy if only one annotator is responsible for each class and we do not have enough reviewers?''
\end{displayquote}
Nati additionally mentioned that the client would not accept the costs of cross-checking results.

For Nati and the QA department, this project involved two challenges:
the first was guessing what the client was expecting from the annotations and which taxonomy would best serve that expectation.
The second consisted in optimizing the performance of annotators to present a competitive offer to the potential client.
Indeed, the Buenos Aires-based company seemed to put much effort into developing better ways of measuring performance and output quality.
In this sense, Nati acknowledged the singularity of the ``gas station project'' as being uncommonly ambiguous compared to the rest of their projects which generally included clear guidelines for the labels.
However, she still saw a good opportunity emerging from the open character of the project:
\begin{displayquote}
``This is where the QA department makes its move and says, okay, we can measure all this.
We try to offer value [...] going into details to see what we can measure and offer the client something they would value because then we also participate in the `farming' process.
If we offer clients valuable QA data, they will probably buy more hours from us.''
\end{displayquote}

\subsection{Project 3: Classifying Faces} \label{sec:proj3}
The third project brings us back to the Bulgarian company (S2). 
It dealt with collections of images depicting people.
All images resembled those commonly found in a mobile phone’s gallery:
several selfies, group pictures of what seemed to be a family, a couple, a child holding a cat.
Eva, the founder of S2, explained that the dataset was intended for a facial recognition model for mobile phones.
The annotations had been commissioned by a local computer vision company.

The first task for the annotators consisted of classifying the faces in the images according to a very concise set of instruction sent via email by the commissioning client:

\begin{displayquote}
(1) For each photo, draw a rectangular bounding box around each face in the photo.

(2) Annotate each such face with the following labels:
Sex: male or female.
Age: baby (0-2 years old), boy or girl (2-16 years old), man or woman (16-65 years old), old man or old woman (65+ years old).
Ethnicity: Caucasian, Chinese, Indian, Japanese, Korean, Latino, Afroamerican.
\end{displayquote}
Additionally, five freely chosen keywords were to be attached to each image.

Founder Eva was in charge of the general quality control.
Apart from her, three annotators completed the team.
Ali, one of the annotators, also managed the project, mostly briefing annotators, tracking the completion of the task, and revising the bounding boxes.
Despite the project's sensitive character, Eva did not have further information about the images' provenance and whether the people depicted were aware their picture would be used in a computer vision product.

Because of the highly subjective character of this project and the specificity of the classes provided, we insistently asked annotators how they were able to differentiate and assign such labels that were, at least to the eye of the researcher on the field, not at all straightforward.
Ali reacted very surprised to this kind of question, almost as if he would not understand our strong interest in this topic:
\begin{displayquote}
``It's not difficult, it's easy! Because all information here [shows the email with the instructions].
You have information.
The woman is between 15 to 65, I think.
The old woman, 65 to more. Old woman and old man.''\\
Interviewer: ``Yeah, but that's what I'm saying, I would have had difficulties telling whether the person in the picture is over 65.''\\
Ali: ``No, no, because you see this picture, you make the zoom, and you see the face [he zooms in and points at the area around the eyes, probably trying to show wrinkles that are hard to recognize as such]. Everything is clear!''
\end{displayquote}
Furthermore, Ali stated that this project was significantly easier to manage than others, given the fact that annotators had not raised any questions or difficulties:
``I think this is a project nobody asked me about,'' he said.
Ali's remarks coincide with the claims of the other annotators involved:
the classification of the people shown in the images in terms of race, age, and sex seemed straightforward to them.
The annotator in charge of keywording also claimed that this task was very easy because the attributes were, in most of the cases, ``pretty obvious.''
When asked what would be the procedure if they were unsure about what labels to assign, Eva, the founder, answered that they would immediately seek the client's opinion:
\begin{displayquote}
``In this case we usually obey everything that they say because you know their interpretations is usually the one that makes sense.''
\end{displayquote}

Later on, Eva mentioned that ``the mobile libraries project'' is one of the most ``controversial'' projects in her company's portfolio. While discussing bias-related issues and how these can affect labels, she also highlighted the importance of raising moral questions around this type of projects and working in solutions for undesirable biases. However, Eva argued that her clients would probably not be interested in investing time or money in these issues.
Similarly, Anna, the intern in charge of conducting an impact assessment at S2, commented on clients' general attitude towards ethical issues related to the commissioned labels:
\begin{displayquote}
``I think even if they knew they should be sensitive or should be a little conscious about these things I think it works in their favor to not be.
It's totally about digital ethics but I feel like it maybe from a company perspective [...] that they would prefer an outsourcing company that doesn't ask too many questions.''
\end{displayquote}
Anna also allocated some responsibility with the annotation companies.
She commented on the difficulty of explaining sensitive categories, such as race and gender, when workers and management have different mother tongues.
In S2, around 98\% of the workers are refugees from the Middle East:
\begin{displayquote}
``Yes, I have observed the [mobile libraries] project ...
I feel a lot of it is not that the company is not aware of these things, but I think it's maybe too complicated to explain to refugees.
I think some of us are lacking the vocabulary that would translate all these nuances.
[...] And I've never heard any of them... any of the refugees ask... I think that's also another factor.
I think it's a combination of a lot of these:
The difficulties to explain it and, maybe, the lack of curiosity or explicit curiosity on their end.''
\end{displayquote}

\subsection{Salient Observations}
\subsubsection{Standardization}
At both annotation companies and in all projects observed, data annotation was performed following the requirements and expectations of commissioning clients.
Guidelines were generally tailored to meet the requirements of the product that would be trained on the annotated datasets, its desired outcome, and its revenue plan.
Instructions and briefings, while providing orientation, aimed at shaping the interpretation of data and, as described by Eva in section \ref{sec:proj1}, ``standardizing everyone's understanding.''
As shown in Projects 1 and 2, quality assurance constituted another decisive instance towards standardization and compliance with clients' expectations.
Encouraged to define what quality means in the context of their company, informants at both locations (S1 and S2) and among the experts (S3) gave more or less different versions of a similar answer:
quality means doing what the client expects.

\subsubsection{Layering}
\begin{figure}[h]
    \centering
    \includegraphics[scale=0.6]{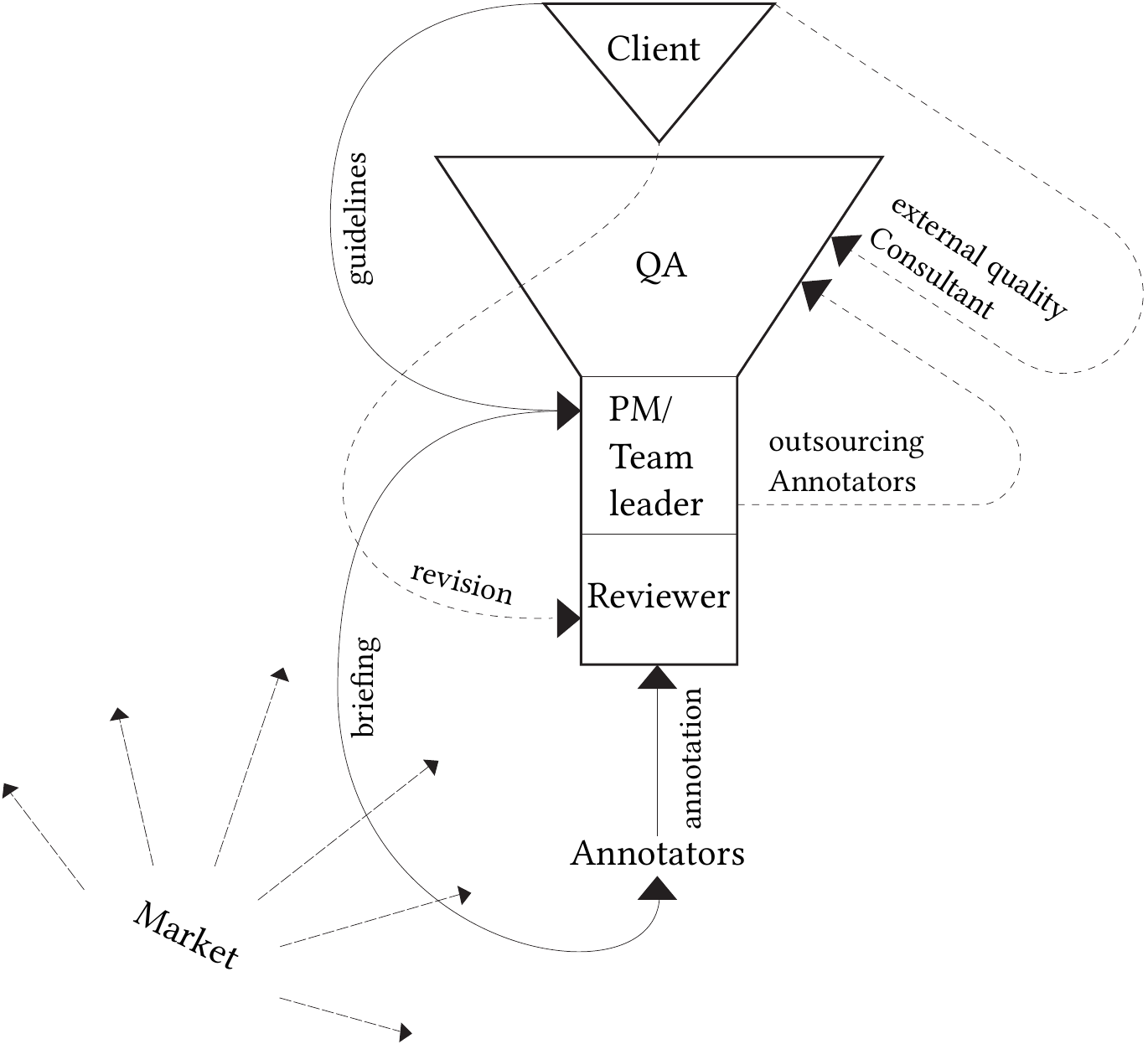}
    \caption{Multiple actors on several layers of classification participate in processes of data annotation. The layers are hierarchical and involve different levels of payment, occupational status, and epistemic authority.}
    \label{fig:annotators}
\end{figure}

As shown in project 2, many roles and departments participate in annotation assignments.
Annotators occupy the lower layer of the hierarchical structure where the actual labeling of data is carried out (see Fig. \ref{fig:annotators}).
In a more or less official way, every company has at least two more layers where control is exercised: reviewers and quality assurance analysts (QA).
In between reviewers and QA, some companies also place team leaders, tech leaders, and project managers.
Finding more layers is possible, depending on the project's and company's size.
As described in Project 2, large corporations sometimes outsource the labeling of the same dataset with different annotation companies.
The results will later be controlled and compared.
Also, important clients often hire external consultants to evaluate the performance of annotation companies independently.
Furthermore, some annotation companies outsource parts of large labeling projects, if they lack the human resources to complete the task. These practices add even more layers to the annotation process.
According to the experts (S3) and practitioners (S4) we interviewed, the layered character of these procedures is not exclusive of S1 and S2 but can be generalized to other annotation companies. 

\subsubsection{Naturalization} \label{sec:naturalization}
Our findings show that the top-down ascription of meanings to data through multi-layered structures were, for the most part, not perceived as an imposition by annotators.
The interviews are abundant in statements such as ``the labels are generally self-evident,'' and ``the work is very straightforward.''

The labels commissioned by clients and instructed by managers seemed to coincide in most cases with annotators' perceptions.
In consequence, labels were hardly ever put under scrutiny or discussed.
Moreover, annotators and managers generally perceived clients to be the ones to know exactly how data was supposed to be labeled since they held decisive information about the product they aimed at developing and the corresponding business plan.
Additionally, in some cases, the image data to be labeled had been directly gathered by the commissioning company, which reinforced the idea that the client would know best how to interpret those images. This was reported by Eva (Founder of S2) in relation to a project involving satellite imagery.
These perceptions contribute to the naturalization of the layers of classification depicted in Fig. \ref{fig:annotators}.
As illustrated by the projects described throughout this section, annotators broadly resolve doubts or ambiguities regarding the labels by asking their superiors.
Both at S1 and S2, we found that the vertical resolution of questions prevailed over horizontal discussions and inter-rater agreement.

\subsubsection{Profit-Orientation}
Annotation companies mostly seek to optimize the speed and homogeneity of annotations to offer reasonable prices in the competitive market of outsourcing services. Several annotators (especially in S2) stated that project deadlines were often so short, that they were difficult to meet. Looking to cope with such a fast pace, workers relied even more on clear guidelines and efficient tools. Several informants at S1 and S2 stated that they found their work easier when clients provided clear instructions, a rather simple platform for the annotations, and a smaller number of classes to label. As shown by the ``gas station project'' (section \ref{sec:proj2}), annotators tended to feel overwhelmed otherwise. In this sense, hierarchical structures did not solely aim at constraining workers' subjectivity but also provided orientation.

As expected from for-profit organizations, commissioning clients and annotation companies are primarily concerned with product and revenue plans.
Moreover, as stated by Eva and Anna in section \ref{sec:proj3}, some annotation companies may perceive a general disinterest of clients regarding the application of ethics-oriented approaches, i.e., transparent documentation and quality control for biased labels.
A similar observation was reported by a QA analyst in S2 and confirmed by the four experts interviewed (S3).
However, this does not mean that detrimental intentions guide clients.
It merely states that ethical approaches involve monetary costs that clients cannot or will not bear.
In short, several informants in S1, S2, S3, and S4 described an environment where market logics and profit-oriented priorities get inscribed in labels, even in projects involving sensitive classifications, as described in section \ref{sec:proj3}.

\section{Discussion}
Our observations show that annotators' subjectivities are, in most of the cases, subdued to interpretations that are hierarchically instructed to them and imposed on data. 
We relate this process to the concept of symbolic power, defined by Pierre Bourdieu~\cite{bourdieu1992} as the authority to impose arbitrary meanings that will appear as legitimate and part of a natural order of things. 
\emph{Arbitrariness} is, in Bourdieu's conception~\cite{bourdieu1990,bourdieu2000,cronin1996}, not a synonym of randomness. It refers to the discretionary character of imposed classifications and their subsumption to the interests of the powerful. 

A \emph{twofold naturalization} in the Bourdieusian sense ~\cite{bourdieu1977}  seems to facilitate the top-down imposition of meaning in data annotation:
\emph{First}, we found that classifications used to ascribe meaning to data are broadly naturalized. 
Annotators mostly perceive the labels instructed by clients and reassured by managers and QA as correct and self-evident.
In a recent investigation, Scheuerman and colleagues present a similar observation, describing how race and gender categories are generally presented as indisputable in image datasets \cite{scheuerman2020}.
In most of the cases observed by us, annotators, managers, and clients do not perceive assigned classifications as arbitrary or imposed. 
Hence, the labels are hardly ever questioned. 
\emph{Second}, we have observed that the epistemic authority of managers and clients is also broadly naturalized by annotators.
They are perceived to know better what labels correctly describe each data point. 
The higher their position, the more accepted and respected their judgments.
Even if annotators or management ever perceive principles of classifications as opposing personal or corporate values, the view persists that ``the one who is paying'' has the right to impose meaning.
This way, clients have the faculty to impose their preferred classifications, just as they have the financial means to pay for labelers to execute that imposition.
As illustrated by the ``gas station project'' in section \ref{sec:proj2}, workers might even feel overwhelmed when clients do not overtly exercise their authority to instruct principles of classification. 
Challenged with making sense of the data themselves, the main rationale becomes ``what would the client want?'' in contrast to ``what is contained in this data?''. 
In this twofold naturalization lies, we argue, the efficacy of interpretations imposed on data:
labels must be naturalized and thus perceived as self-evident if actors are to misrecognize the arbitrariness of their imposition~\cite{bourdieu1977}. 

As shown by our findings, the standardization of annotation practices and labels is assured throughout several layers of classification and control. 
The positions are depicted in Figure \ref{fig:annotators} as hierarchical layers positioned one above the other because they involve different levels of responsibility, payment, and occupational status.
The number of layers, actors, and iterations involved hinders the identification of specific responsibilities. 
Moreover, no information regarding the actors involved and the criteria behind data-related decisions is registered. 
Annotation steps and iterations remain broadly undocumented.  
Accountability is diluted in these widespread practices. 
A problematic implication is that the multi-layered standardization process is hardly ever oriented towards social responsibility and usually responds to economic interests only~\cite{kazimzade2020}. 
There is no intention, however, to imply here that standardization is fundamentally harmful or that detrimental intentions lead the actors involved. 
We rather aim at showing how power structures can be stabilized through imposed standards \cite{bowker1999} and argue that standardization can be dangerous if it is solely guided by profit maximization.

In this sense, we argue that the discussion on workers' subjectivity and personal values around data annotation should not let us researchers forget that datasets are generally created as part of large industrial structures, subject to market vicissitudes, and deeply intertwined with naturalized capitalistic interests. The challenge here is ``to explicate the assumptions, concepts, values, and methods that today seem commonplace''~\cite{blomberg2015} in this (and other) forms of service.

The \emph{main contribution} of our investigation is the introduction of a power-oriented perspective to discuss the dynamics of imposition and naturalization inscribed in the classification, sorting, and labeling of data. 
Through this lens, we shed light on power imbalances informing annotation practices and shaping datasets at their origins. 
Our main argument is that power asymmetries inherent to capitalistic labor and service relationships have a fundamental effect on annotations.
They are at the core of the interpretation of data and profoundly shape datasets and computer vision products.

There are at least two close-connected reasons why imposition and naturalization in the context of data creation are socially relevant and, in a way, different from power imbalances enacted through work practices in other settings: 
\emph{First}, data practices involve particular ethical concerns because assumptions and values that inform data can potentially have devastating effects for individuals and communities~\cite{eubanks2018,oneil2017}. 
Algorithms trained on data that reproduces racists, sexist, or classist classifications can reinforce discriminatory visions~\cite{noble2018} ``by suggesting that historically disadvantaged groups actually deserve less favorable treatment''~\cite{barocas2016}. Moreover, data about human behavior is increasingly sold for profit \cite{zuboff2019}, which could result in surveillance~\cite{zuboff2019} and exploitation~\cite{couldry2019}.
\emph{Second}, data-related decisions define possibilities for action, by making certain aspects of reality visible in datasets, while excluding others~\cite{pine2015,bowker2000b}.
This is relevant for state management and policy, i.e., to pinpoint places where intervention or allocation of resources is needed. However, the tendency of classification practices towards the erasure of residual categories~\cite{bowker1999} can cause tension and even be harmful for individuals who remain unseen or misclassified by decision-making systems ~\cite{scheuerman2019, buolamwini2018}.

\subsection{Implications for Practitioners}
While annotation companies and their clients may or may not be aware that they are actively shaping data, the opacity surrounding embedded interests and preconceptions \cite{scheuerman2020} is a significant threat to fairness, transparency, accountability, and explainability. Therefore, it is important that practitioners, i.e., corporations commissioning datasets and management at annotation companies, take steps to reflect, document, and communicate their subjective choices~\cite{muller2019,passi2017,passi2019,gebru2018, scheuerman2020}. Promoting the intelligibility of datasets is fundamental because they play a key role in the training and evaluation of ML systems.  Understanding datasets' origin, purpose, and characteristics can help better understand the behavior of models and uncover broad ethical issues~\cite{wortmanvaughan2020}.

Recent research work has highlighted the importance of structured disclosure documents that should accompany datasets~\cite{gebru2018,madaio2020,geiger2020,bender2018,holland2018,wortmanvaughan2020}.
Fortunately, the machine learning research community has begun to promote similar reflexive practices:
Following Pineau's suggestion~\cite{pineau2020}, authors of NeurIPS and ICML conference are now requested to include a reproducibility checklist which encourages “a complete description of the data collection process, such as instructions to annotators and methods for quality control” if a new dataset is used in a paper.
NeurIPS further requires authors to disclose funding and competing interests.
They are also asked to discuss “the potential broader impact of their work, including its ethical aspects and future societal consequences.”
These conferences are highly influential for ML practitioners and facilitate the adoption of the latest machine learning capabilities.
It is certainly our hope that they will also inspire them to adopt such reasonable best practices and to engage in reflexive documentation.

In line with previous literature~\cite{gebru2018,kazimzade2020,bender2018,seck2018,wortmanvaughan2020}, we advocate for the documentation of purpose, composition, and intention of datasets.
Moreover, the structures, decisions, actors, and frameworks which shape data annotation should be made explicit~\cite{geiger2020,scheuerman2020}.
We furthermore propose orienting documentation towards a reflexion of power dynamics. 
D'Ignazio and Klein \cite{dignazio2020} propose asking \emph{who questions} to examine how power operates in data science. 
In this vein, we propose that disclosure documents include answers to questions such as:
Whose product do the annotations serve, and how?
Whose rationale is behind the taxonomies that were applied to data? Who resolved discrepancies in the annotation process?
Who decided if labels were correctly allocated?

We argue that the annotation process already begins as clients transform their needs and expectations into annotation instructions.
Therefore, the responsibility for documenting should not be solely placed with annotators but should be seen as a collaborative project involving annotation companies and commissioning clients.
Given the hierarchical structures and power imbalances described in this paper, we find it extremely important that clients keep a record of the instructions that were given to annotators, the platforms on which annotations were performed, and the reasons for that platform choice, as well as the procedure employed for solving ambiguities, creating homogeneity, and establishing inter-annotator agreements.
Extending dataset factsheets with a power-aware perspective could make power asymmetries visible and raise awareness about meaning impositions and naturalization.
Yet, it is vital that documentation checklists are not prescriptive and produced exclusively in the vacuum of academia~\cite{gebru2018}.
Instead, disclosure documents should be developed in an open and democratic exchange with annotation companies and their clients to accommodate real-world needs and scenarios~\cite{madaio2020}.

Annotation companies and their clients might be reluctant to implement such a time-consuming documentation process.
Moreover, they may regard some of the information as trade secrets, especially if it involves details about the intended product or if the structuring of the annotation process is considered a strategic advantage.
We argue that allocating resources for documentation could nevertheless bring three pay-offs for organizations:

\emph{The first benefit} is that proper documentation can foster deliberative accountability~\cite{passi2018} and improve inter-organizational traceability, for instance, between annotation companies and clients.
In addition, transparent documentation can help address the problematic dilution of accountability as a result of various actors and layers in the annotation process.
In the context of this service relationship, accountability involves not only specific individuals but also organizations and includes factors such as organizational routines and processes of value co-creation~\cite{kimbell2017}.
Given the power imbalances that are inherent to this relationship~\cite{blomberg2015}, annotation companies could be motivated to keep track of decisions and procedures in the event of discrepancies with clients.  

\emph{The second benefit} is that documentation can facilitate compliance with regulations such as the GDPR and especially the ``Right to Explanation''~\cite{passi2018}.
Serving as an external motivation, legal frameworks and regulations urge companies to put transparency as well as societal and ethical consequences of their products and services above the rationale of profit-maximization~\cite{kazimzade2020}.
If there is no legal incentive and companies perceive transparency as coming at the cost of profit-oriented goals (as shown in our data), independently created transparency certifications and quality seals for datasets may provide an additional incentive given the momentum created around FATE AI.

\emph{The third benefit} is that documentation may create a long term business asset because knowledge about practical data work is made explicit and persistent.
Without documentation, such knowledge is often confined to workers with the ``craftsmanship'' to make situated and discretionary decisions~\cite{passi2017}, bearing the risk of knowledge loss due to worker flow or lack of traceability. 
At the same time, documentation can have analytical value, improve communication in interdisciplinary teams, and ease comprehension ``for people with diverse backgrounds and expertise''~\cite{muller2019}.

\subsection{Implications for (CSCW) Researchers}
Our research highlights the relation between human intervention and hierarchical structures in processes of data creation.
It shows that power imbalances not only translate into asymmetrical labor conditions but also concretely shape labels and data.
We firmly believe that researchers studying socio-technical systems in general, and data practices, in particular, could benefit from including a similar, power-aware perspective in their analysis. 
Such a perspective would primarily aim at making asymmetrical relations visible. 
Making power visible means exposing naturalized imbalances that get inscribed in datasets and systems~\cite{dignazio2020}. 

We propose four (interconnected) reasons for integrating such a perspective into research:

\emph{First}, this perspective could contribute to making work visible~\cite{gray2019,star1999,dignazio2020}.
Especially in the case of machine learning systems where the enthusiasm of technologists tends to render human work invisible~\cite{gray2019}, research should emphasize the value of the human labor that makes automation possible.
Furthermore, making ``humans behind the machines''~\cite{klinger2018} visible could help contest any pretension of calculative neutrality attributed to automated systems.

\emph{Second}, this paper argues that power relationships inscribed in datasets are as problematic as individual subjectivities.
A power-oriented perspective allows researchers to ``shift the gaze upwards''~\cite{barabas2020} and move beyond a simplistic view of individual behaviors and interpretations that, in many cases, could end up allocating responsibilities with workers exclusively.
A view into coroporate structures and market demands can offer a broader perspective to this line of research.

\emph{Third}, the investigation of organizational routines and hierarchies could help researchers approach the real-world practice of data work~\cite{passi2018}, develop context-situated recommendations, and assess their applicability in corporate scenarios.
This could help establish open and democratic discussions between researchers and practitioners regarding the conception of solutions for undesired data-related issues~\cite{gebru2018,madaio2020}.

\emph{Finally}, rigorous reflexion and documentation of power dynamics is not only advisable for practitioners working with data but is also fundamental for researchers investigating those work practices.
Acknowledging that, just like data, theories are not discovered, but they are co-constructed by researchers and participants~\cite{thornberg2012} is a significant step in this direction.
Throughout this investigation, the constructivist variation of grounded theory~\cite{charmaz2006} has constituted a fantastic tool to methodically reflect on the researchers' perspectives, interpretations, and position.

\subsection{Limitations and Future Work}
This paper has focused on the annotation of image data for machine learning as performed within impact sourcing companies.
While our current results are bound to this context, the framework presented here could inspire further (comparative) research involving diverse actors in other annotation settings, such as crowdsourcing platforms.

\section{Conclusion}
This paper has presented a constructivist grounded theory investigation of the sensemaking of data as performed by data annotators.
Based on several weeks of fieldwork at two companies and interviews with annotators, managers, and computer vision practitioners, we have described the structures and standards that influence the classification and labeling of data. We aimed at contesting the supposed neutrality of data systems by setting the spotlight on the power dynamics that inform data creation.

We found that workers' subjectivity is structurally constrained and profoundly shaped by classifications imposed by actors above annotators' station. Briefings, annotation guidelines, and quality control all aim at meeting the demands of clients and the market.
We have argued that the creation of datasets follows the logics of cost effectiveness, optimization of workers' output, and standardization of labels, often to the expense of ethical considerations.

We have observed the presence of multiple instances of classification, with diverse actors among several hierarchical layers that are related to the possession of capital.   
We have argued that the many layers, actors, and iterations involved contribute to the imposition of meaning and, finally, to the dilution of responsibilities and accountability for the possible harms caused by arbitrary labels.
Furthermore, our findings have shown that workers naturalize the imposed classifications as well as the epistemic authority of those actors higher in the hierarchy.
Our observations indicate that power asymmetries, which are inherent to labor relations and to the service relationship between annotation companies and their clients, fundamentally shape labels, datasets, and systems.

We have furthermgore discussed implications for practitioners and researchers and advocated for the adoption of a power-aware perspective to document actors and rationale behind the meanings assigned to data in annotation work.
Finally, we have emphasized the importance of adopting a similar power-aware perspective in the CSCW research agenda, not only as a possible focus for future work but also as a tool for reflecting on researchers' own position and power.

\section{Acknowledgements}
Funded by the German Federal Ministry of Education and Research (BMBF) – Nr 16DII113. We would like to acknowledge the individuals and companies participating in this study: we dearly thank them for their openness!  Special thanks to Philipp Wei{\ss} for his support whenever we struggled with formatting tables in Overleaf. We wish to thank our anonymous reviewers for their feedback, and Enrico Costanza, Walter S. Lasecki, Leon Sixt, and Alex Hanna for valuable comments on earlier versions of this work.

\bibliographystyle{ACM-Reference-Format}
\bibliography{references}

\end{document}